# Mapping microstructure to shock-induced temperature fields using deep learning

Chunyu Li [*], Juan Carlos Verduzco [*], Brian H. Lee, Robert J. Appleton, and Alejandro Strachan

School of Materials Engineering, Purdue University

West Lafayette, Indiana 47907 USA

The response of materials to dynamical, or shock, loading is important to planetary science, aerospace engineering, and energetic materials. Thermal-activated processes, including chemical reactions and phase transitions, are significantly accelerated by the localization of the energy deposited into hotspots. These results from the interaction of a supersonic wave with the materials' microstructure and are governed by complex, coupled processes, including the collapse of porosity, interfacial friction, and localized plastic deformation. These mechanisms are not fully understood and today we lack predictive models to, for example, predict the shock to detonation transition from chemistry and microstructure alone. We demonstrate that deep learning techniques can be used to predict the resulting shock-induced temperature fields in complex composite materials obtained from large-scale molecular dynamics simulations with the initial microstructure as the only input. The accuracy of the Microstructure-Informed Shock-induced Temperature net (MISTnet) model is higher than the current state of the art at a fraction of the computation cost.

## INTRODUCTION

The chemical and physical processes launched by a shockwave as it travels supersonically through a material are important in planetary sciences [1], microparticle impact [2], synthesis of new materials [3], and the initiation of detonations [4, 5]. Thermally activated processes triggered by dynamical loading can be accelerated significantly by the localization of energy into hotspots [6], which form by the interaction of the supersonic wave and the materials' microstructure or by shock-induced defects. A striking consequence of the importance of hotspots is the near impossibility of detonating many single-crystal molecular explosives. Multiple processes lead to energy localization, most prominently the collapse of porosity (which involves severe plastic deformation, local amorphization, and jetting), interfacial friction, the propagation of cracks, and localized plastic deformation. Non-trivial challenges in the prediction of shock-induced hotspot formation are the fact that these processes are strongly coupled and occur under extreme conditions of temperature, pressure, and strain rate. Given the disparate length and time scales associated with the various mechanisms of energy localization and microstructural features, it is not surprising that no single model can predict shock-induced hotspot formation without significant approximations [7]. In this paper, we show that deep learning can map the initial microstructure of a complex composite material consisting of crystalline grains and a polymer binder to the resulting temperature field, including hotspots. Importantly, the training and testing data were obtained from atomistic simulations that explicitly capture all the above mechanics many of which are sub-grid to the input and output fields of the model.

---

[*] Equal contributions

Continuum models capable of capturing the microstructure of the materials of interest necessarily approximate many of the mechanisms of hotspot formation, including localized plastic deformation and pore collapse. Hotspots are often considered sub-scale and introduced in a statistical manner [8,9]. In contrast, atomistic simulations that describe molecular phenomena explicitly cannot reach the characteristic scales of important microstructural features. MD has been extensively used to study shock-induced hotspot formation, including the collapse of porosity, shear, friction, and localized plastic deformation. These studies have provided valuable insight into the mechanisms undergirding the formation of hotspots and the relative potency of various processes. Yet, we still lack models to predict the hotspot temperature and size resulting from the various energy localization mechanisms, except for simple processes and simple geometries [10]. Recent efforts have compared the temperature fields obtained from MD simulations with multi-physics, continuum-level models [11, 12, 13]. These comparisons revealed the progress in continuum modeling but also point to their intrinsic limitations and the need for nano-meter scale resolution with the consequent computational cost. In addition to physics-based approaches, deep learning has been used to model the mesoscale thermo-mechanical response of materials under shock loading with accuracy comparable to physics-based simulations but at a fraction of the computational cost [14].

To address this current gap, we explore whether deep learning [15] can be used to develop computationally inexpensive models capable of mapping complex microstructures and defects to shock-induced temperature fields from explicit MD simulations. Importantly, given the significant computational cost of such MD simulations (e.g. a single shock simulation on a system measuring 100 nm in length with 1200 $nm^2$ cross-sectional area requires approximately 28,000 CPU hours on AMD Epyc 7662 cores), the models should be developed using a relatively small number of simulations. Inspired by work on field-to-field mapping using deep learning [16,17], we reduce the initial atomic structure into three scalar fields and the final temperature into a single field. We find that a dozen independent simulations enable the training of a U-Net-type [17] deep neural network capable of predicting the temperature field associated with shock loading of microstructures not seen during training, including the location and temperature distribution of hotspots. The Microstructure-Informed Shock-induced Temperature net (MISTnet) model, see Figure 1, shows comparable or better accuracy than physics-based and machine-learning models at a fraction of the computational cost. Convolutional neural networks, including U-nets have been used in materials science, with great success, for image classification and segmentation, regression, and field to field predictions, and mesoscale modeling. Unlike prior efforts, we demonstrate the use of deep learning to by-pass an explicit physics-based simulation that resolves physics that are sub-scale to MISTnet and are described via a particle-based Lagrangian framework which is mapped into a Eulerian framework that deforms affinely with the material.

## RESULTS AND DISCUSSION

### Explicit simulations of hotspot formation in shocked composites

To explicitly capture all mechanisms of hotspot formation in realistic structures, we performed all-atom MD simulations of shock propagation on complex microstructures consisting of polymer-bonded molecular crystals. Such microstructures are common in energetic materials and particle-reinforced polymer composites and include interfaces between different materials, anisotropic crystals, and porosity. The explicit description of every atom in the systems, see Figure 1(a), enables the explicit simulation of all the processes expected to result in hotspot formation, the only approximation being the description of atomic interactions and the use of classical mechanics to describe atomic dynamics. Pore collapse, crack propagation, localized plasticity, and interfacial friction are described without further approximations or modeling. The composite systems consist of 1,3,5-Trinitroperhydro-1,3,5-triazine (RDX) grains and polystyrene (PS) binder; they were built using our in-house PBXgen code [18]. Atomic interactions are described with extensively validated force fields [19, 20] that provide an accurate description of the thermo-mechanics of these materials [21, 22, 23]. The neglect of chemical reactions is another approximation, but the focus of this paper is on the prediction of hotspot formation right after the passage of the shock (1 ps after the shock front) when hotspots are fully formed but chemistry is negligible.

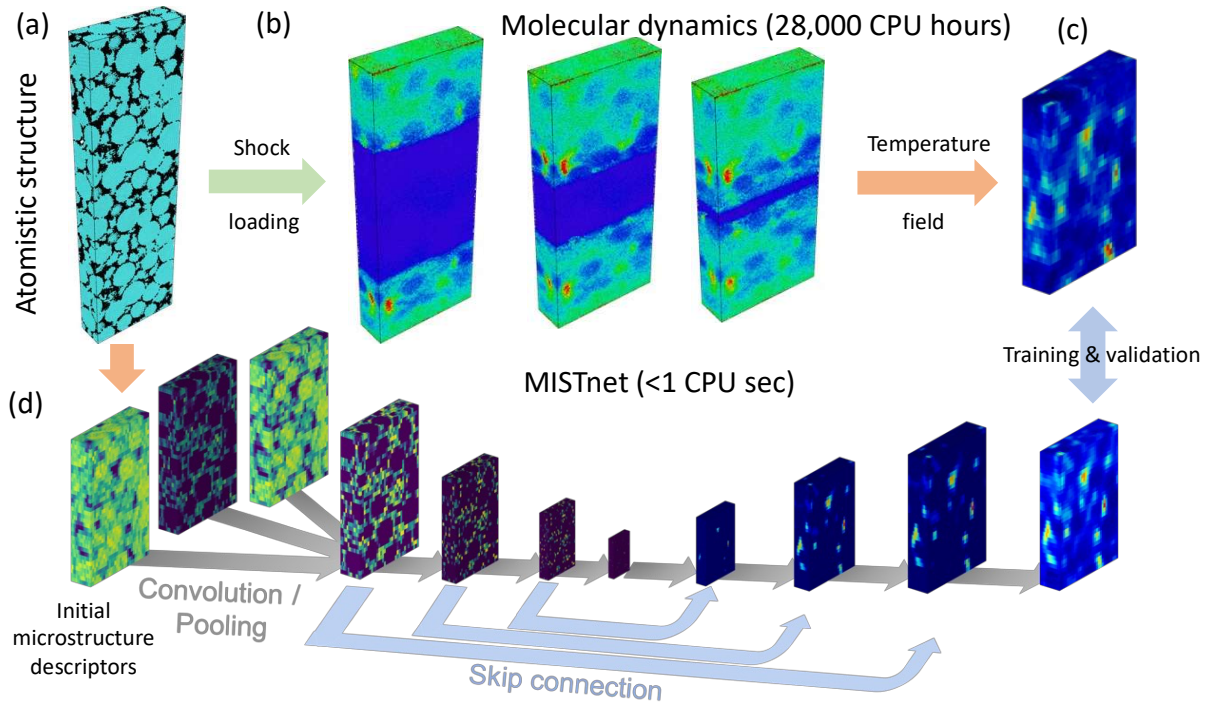

**Figure 1** Schematic representation of our approach to learning shock-induced temperature fields. (a) Initial atomistic structure of composite; (b) Large-scale MD simulation of shock propagation; (c) Shock induced temperature field from MD; (d) MISTnet architecture and input and output fields.

**MD simulations for training and testing.** We built and shocked 11 independent RDX/PS systems with two different PS wgt. %, spherical or facetted RDX crystals, and two cell sizes. The particle diameters follow a bimodal distribution, with maxima at 8 and 17 nm, scaled down from typical experimental data in energetic materials [24]. We built four systems containing approximately 30 million atoms (with cell dimensions of approximately 84 nm x 18 nm x 220 nm, these larger systems will be denoted L1 to L4) and seven systems with 10 million atoms (approximately 60 nm x 18 nm x 100 nm) denoted S1 to S7. Details of the various systems are included in Table 1.

**Table 1.** Details of the atomistic PBX systems (Tr, V, and Tst represent training, validation, and testing datasets)

| PBX | RDX crystal particle | | PS content (weight%) | Dimension (nm) Lx, Ly, Lz | Density (g/cm$^3$) | Datasets |
|---|---|---|---|---|---|---|
| | Shape | Orientation | | | | |
| L1 (Tr) | Spherical | Oriented | 14.7% | 84.24, 18.17, 222.24 | 1.469 | 16 |
| L2 (Tr) | | | | 83.16, 17.60, 220.02 | 1.460 | 16 |
| L3 (Tr) | | Random | 9.5% | 82.34, 17.84, 218.75 | 1.464 | 16 |
| L4 (Tr) | Facetted | | | 87.51, 18.06, 238.68 | 1.466 | 16 |
| S1 (V) | Spherical | Oriented | 14.7% | 61.73, 18.17, 102.65 | 1.482 | 4 |
| S2 (Tst) | | | | 61.75, 17.61, 97.91 | 1.499 | 4 |
| S3 (Tst) | | Random | 9.5% | 61.43, 17.69, 98.43 | 1.492 | 4 |
| S4 (Tst) | Facetted | | | 63.58, 17.94, 109.65 | 1.491 | 4 |
| S5 (V) | Spherical | | 14.7% | 57.14, 18.53, 108.95 | 1.471 | 4 |
| S6 (V) | | | | 59.63, 18.23, 106.56 | 1.464 | 4 |
| S7 (V) | | | | 59.88, 18.46, 102.65 | 1.493 | 4 |

Each PBX system was subjected to shock loading using adiabatic MD via a convergent shock approach [25]. This setup generates two shocks at the cell z boundary that propagate toward the center of the simulation box, see Fig. 1(b). Thus, to maximize the information generated, the PBX systems are replicated along the shock direction prior to dynamical loading. This results in shocks propagating on each microstructure in opposite directions. The particle velocity was 2.5 km/s, which results in a shock velocity of approximately 6.7 km/s.

### Input and output fields and deep convolutional neural networks

Our goal is to develop predictive models that can relate the initial structure to the resulting temperature field following shock loading. Prior efforts involving deep learning of field variables in the physical sciences and engineering have focused on continuum models or measurements [26, 27]. In contrast, we are interested in predicting the output of an explicit, particle-based simulation with chaotic dynamics. Thus, the first step in model development is encoding the inputs and outputs into field variables. This is not uncommon for output variables like temperature. We computed the center-of-mass temperature of each RDX molecule and polymer monomer and mapped them into an orthorhombic grid to obtain a field of local temperature after the passage of the shock. The input to the atomistic simulation is a set of (tens of millions) atomic coordinates and velocities that need to be converted into field variables. We choose three local properties to characterize the initial microstructure: i) total density, ii) density of RDX, and iii) the product of RDX and PS densities to highlight interfaces, see Fig. 1(d). We hypothesized that these descriptors could capture hotspot mechanisms like pore collapse and interfacial friction. The input fields are mapped over a grid consisting of 32 x 16 x 32 bins (each bin approximately 1.9 nm x 1.1 nm x 2.8 nm) for the S1 to S7 systems. The output field (local temperature of shocked PBX structure) is mapped into the same number of bins with length along the shock direction reduced to 1.7 nm due to shock compression. For systems L1 to L4, we used similarly sized bins but larger grids of 44 x 16 x 80. In total, we prepared 92 datasets consisting of three input fields and one output field: 28 from the 7 small PBX systems S1 to S7 (considering the two shocks per simulation and data augmentation by subjecting each input-output pair to a mirror operation on the plane containing the shock direction and the longer transverse direction), and 64 from the L1 to L4 (two shocks per simulation, each divided into four regions of size matching the small cells, top-left, top-right, lower-left, lower-right, plus mirror augmentation). Note that datasets obtained from the L1-L4 simulations have partial overlap.

Our MD data was split into training and testing sets. As mentioned previously, each simulation was included a second time, mirrored, as a data augmentation strategy. The complete set of simulations was split into training (~70%, marked Tr in Table 1), validation (~17%, marked as V), and testing (~13%, marked as Tst). To avoid data leakage, all the datasets originating from each simulation (including the multiple datasets from the large systems and mirrored images) were sorted together into a training, validation, or testing set. The validation set was sequestered and only used to control the early stopping criteria needed to prevent overfitting.

Deep neural networks are known to be universal approximators [28]. These models can model complex non-linear functions between inputs and outputs by creating connections through internal parameters that can be considered as features of the input data and activation functions to act on these parameters. Convolutional Neural Networks (CNNs) have been particularly successful in image-processing tasks like classification and image semantic segmentation [29,30] with applications in generative modeling [31] and medical imaging [32]. A commonly implemented technique for automatic image segmentation is based on an encoder-decoder architecture called U-Net, see Figure 1(d). First proposed by Ronneberger et al. [17], U-Nets use layers of convolution and pooling to extract and process information at various scales and a "skip-connection" approach to convey information from the encoder to the decoder stage, thus improving feature localization. In materials science, U-nets have been used to analyze 2D images and identify precipitates in Ni-superalloys [33], recognize nanoparticles [27], and locate atomic defects to characterize degradation [34].

To map the initial microstructure fields to the shock temperature field, MISTnet uses 3x3x3 volumetric convolution blocks (a domain approximately 6 nm on the side), max pooling, and ReLU activation functions. Batch normalization operations were applied before pooling layers to regularize the model variance. Pooling layers were used to down-sample feature maps along their spatial dimensions by taking the maximum value over a window for each channel. By pooling these layers, information about regions surrounding each microstructural feature can be communicated in downstream operations. A maximum value pooling was selected as it is the deviation from the average microstructure that is expected to result in higher temperatures. The pooling stride was selected to avoid overlapping windows. To replicate the periodic boundary conditions on the simulation, a periodic padding function was selected along the spatial directions that do not match the shock direction. This function applies to the convolutional layers before the first pooling operation.

The model was trained using the Adam [35] optimizer with a custom loss function based on the mean-square error that increases the contribution of voxels with target temperatures over a threshold of 1800 K by a factor of five to best capture hotspots. The learning rate decreases from 0.0005 to 0.0001 after 500 epochs and we use an early stopping criterion based on the validation loss, which stops the model if no improvement over 0.0005 is observed in 200 epochs, restoring the best weights in the model.

## Model accuracy

Figure 2 illustrates the performance of MISTnet for test set S2. Visual inspection of the predicted and MD temperature fields clearly shows the model's ability to predict key features of the temperature field arising from the interaction between the microstructure and the shock. The location, size, and intensity of the hotspots are clearly captured. MISTnet not only predicts the hotspots associated with porosity (see red, orange, and black arrows) but also captures non-trivial features. For example, the model learned that voids elongated along the direction of the shock (orange) result in higher temperatures than those elongated in the transverse directions (black) [36]. Beyond porosity, interfacial processes like friction and acoustic mismatch are known to result in energy localization, and our model is also capable of capturing those; an example is highlighted using green arrows.

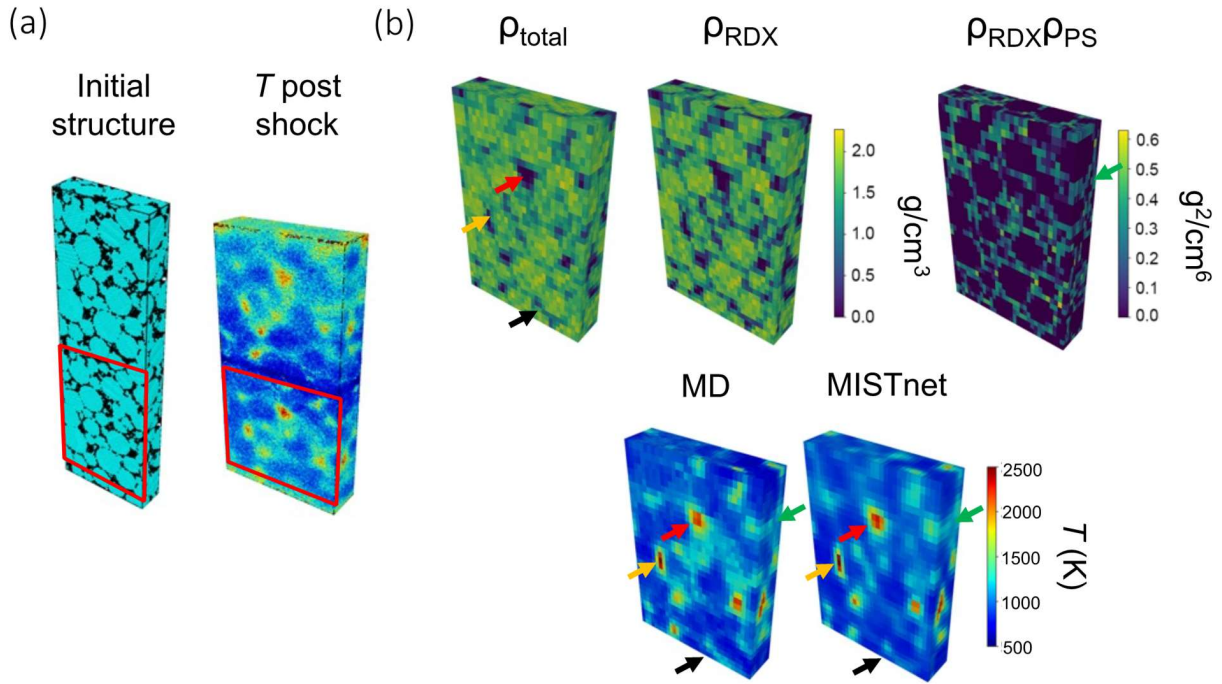

**Figure 2** (a) Initial atomistic structure and shock-induced temperature. (b) Binned data of input data ($\rho_{total}$, $\rho_{RDX}$, $\rho_{RDX}\rho_{PS}$) and outputs comparing the MD and MISTnet results temperature fields.

To quantitatively analyze the accuracy of the field predictions, we performed a series of comparisons with varying levels of granularity. At the finest scale, we compare the temperatures obtained from MD and MISTnet at the bin level. For this analysis, we apply a Gaussian smoothing filter with characteristic length equal to a bin (~2 nm) to the temperatures to account for the stochastic nature of MD that leads fluctuations of local temperature. The resulting parity plot between temperature predictions vs. MD results for test set S2, Figure 3(a), demonstrates the accuracy of MISTnet. A comparison of the raw, unfiltered, data is included in Fig. S1 of the SI. Furthermore, as the criticality of hotspots is known to be a function of both the hotspot temperature and size, we analyzed the distribution of volume of the material with the temperature exceeding a value T. Figure 3(b) shows this distribution as T vs. volume. Again, we observe a very good agreement between MD and MISTnet. Figure 3 shows that MISTnet tends to underestimate the highest temperatures but these command relatively small volumes.

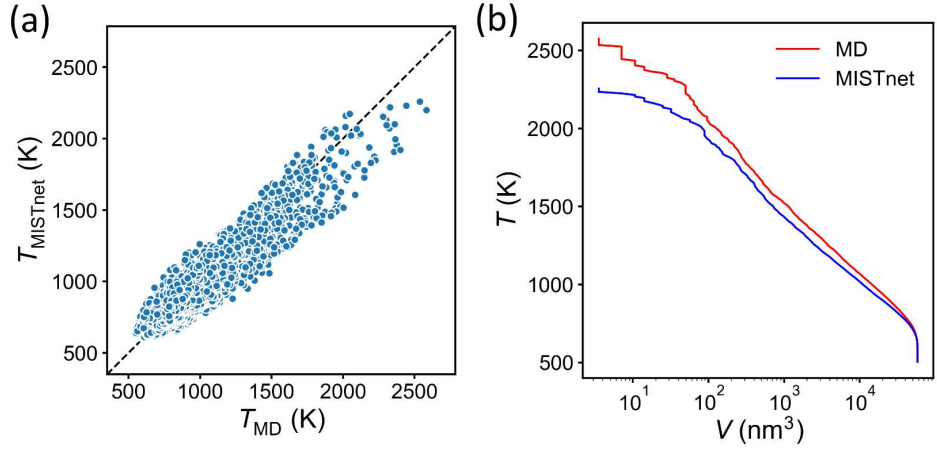

**Figure 3**. Comparison of MD and MISTnet results. Parity plot (a) and cumulative volume vs. T; (b) for all bins of test dataset from the S2 system. Gaussian smoothing filter is applied to the raw temperatures obtained from MD and MISTnet

To compare our model with prior work, we calculated the average temperature and volume of the hotspots from MD and MISTnet for all cases simulated, see Figure 4. Here, we define hotspot as those bins with temperature exceeding 1800 K and analysis using hotspot cutoff temperatures of 1600 K and 2000 K show similar results, see Fig. S2 in the SI. Across all training, validation, and test sets, we observe good agreement between MD and MISTnet. The RMS error in temperature is $\Delta \bar{T}_{\text{hotspot}} \sim 50$ K and that in volume $\Delta \bar{V}_{\text{hotspot}} \sim 200$ nm$^3$, see Table 1. When normalized by the average values, the errors in hotspot temperature and volume are 2.3 % and 15%, respectively. While direct comparison cannot be made due to the difference in shock conditions, the errors in $\Delta \bar{T}_{\text{hotspot}}$ are approximately factors of 7 times smaller than those reported for the Physics-Aware Recurrent Convolutional Neural Networks (PARC) model [14].

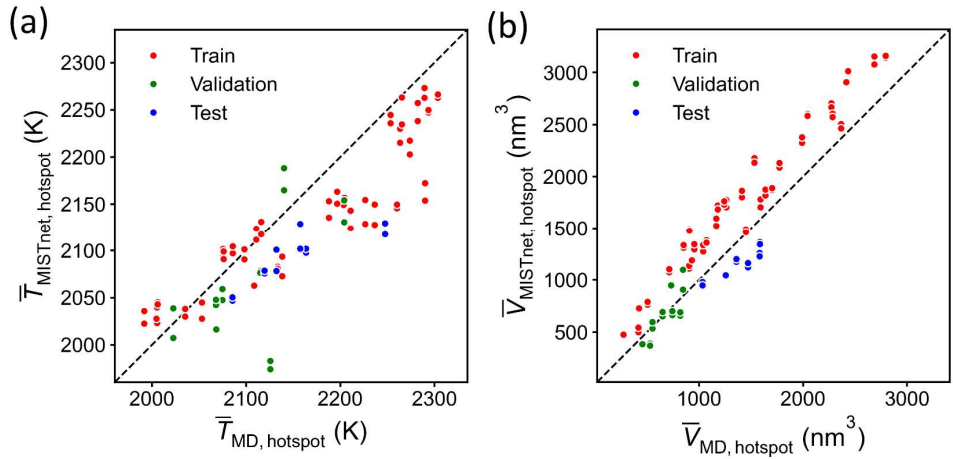

**Figure 4**. Comparison of hotspots obtained from MD and MISTnet. Average hotspot temperature (a) and volume (b) for all cases.

**Table 2** Error analysis of the MISTnet predictions

| RMSE | T field (K) | $\Delta \bar{T}_{hotspot}(K)$ | $\Delta \bar{V}_{hotspot}(nm^3)$ |
|---|---|---|---|
| Train | 72.57 | 41.96 | 350.84 |
| Validation | 91.57 | 47.37 | 107.07 |
| Test | 97.62 | 58.38 | 224.16 |

While the measures above quantify the overall temperature distribution and average hotspot properties, the criticality of a hotspot depends on their individual characteristics [37]. Therefore, we developed a novel one-by-one comparison of predicted vs. MD hotspots. We identify individual hotspots via a cluster analysis over grid points with local temperatures above 1800 K, using a distance cutoff slightly longer than the body diagonal of a single bin. After the hotspots are identified we use an overlap criterion to map MD and MISTnet hotspots, see Section S-3 of the SI for the detailed procedure. Figures 5(a) and 5(b) show the spatial distribution of the MD and MISTnet hotspots, respectively. We use the same color to indicate matching hotspots predicted by both MD and MISTnet model and black for hotspots only seen in the MD or MISTnet. As indicated above, we observe remarkable agreement in the location and shape of hotspots. To further quantify the accuracy of the mode, Figure 5(c) compares the temperature distribution vs. characteristic length (cube root of total volume) of individual hotspots for the MD (left-sided violins) and MISTnet (ride-sided violins); identical colors mark matching hotspots. We find several hotspots that present in both the MD and MISTnet that have comparable size and temperature distribution. Arrows connect matching hotspots. MISTnet slightly underestimates the size (by approximately 10%) and highest temperatures of the largest hotspot. Only a few small hotspots appear either in the MD or in the MISTnet predictions.

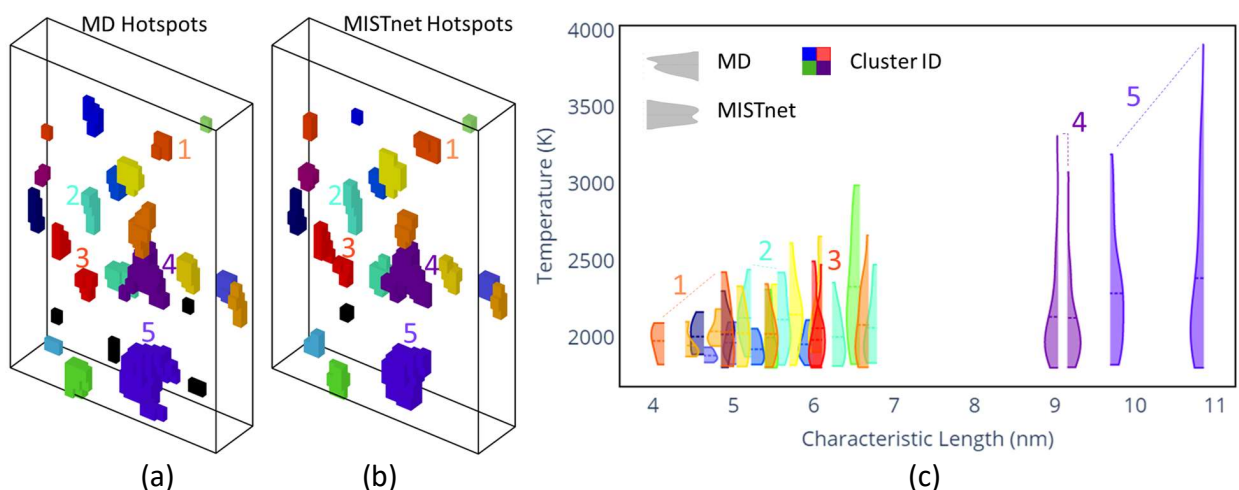

(a)      (b)      (c)

**Figure 5**: Cluster analysis for a typical testing data set from the S2 system. The hotspot from MD (a) and predicted by the MISTnet (b) are marked with identical colors when they are matched and black when they are only present in a single field. (c) The temperature distributions of the matched hotspots as a function of characteristic size (left-sided distributions represent MD results, and the right-handed distributions MISTnet), selected hotspots are numbered for easy identification.

## CONCLUSIONS AND OUTLOOK

This paper introduced a deep learning model, MISTnet, capable of predicting the temperature field resulting from the shock loading of a composite material with microstructure as the only input. The training, validation, and testing data are obtained from multi-million-atom MD simulations that explicitly describe all the mechanisms underlying the formation of hotspots. We find that three input fields (total density, density of one of two materials in the composite, and the product of densities to highlight interfaces) and about a dozen atomistic simulations enable MISTnet to map the initial, pre-shock, microstructure to shocked temperature field. Once trained, MISTnet can predict temperatures of unseen microstructures with a computational cost of $10^{-4}$ that of an MD simulation. MIST-net can predict hotspots originating from the collapse of porosity and localized shear and learn subtle effects like the effect of pore orientation. As an additional test to stress-test the model, we applied it to predict the hotspot resulting from a single cylindrical pore, 20 nm to 50 nm in diameter, in a pure RDX crystal. This microstructure, with no polymer, is very different from those of the composites used in training, validation, and testing and contains a pore much larger than anything MISTnet has ever seen. Remarkably, MISTnet predicts a crescent shape hotspot with a very reasonable temperature distribution, see Section S-4 in the SI. The complexity of the microstructures and accuracy of the temperature field predictions go beyond what has been attempted to date either with ML or physics-based models.

MISTnet temperature fields can be used as input to models of thermo-mechanics and chemical kinetics to simulate subsequent processes. We envision MISTnet as a key component in future continuum-scale models, providing sub-scale microstructure-aware information, for shock-induced phenomena, from materials processing to detonation. In this first effort, MISTnet was trained using a single particle velocity, limiting its use. While it should be possible to train a series of independent models for various shock strengths, we believe a parametric approach adding an additional input indicating the particle velocity will be very valuable.

## METHODS

**Atomistic simulation on PBX systems**

Large scale atomistic MD simulations on PBX systems are conducted by using the open-source software LAMMPS [38]. The non-reactive molecular force field for PS is from Dreiding [20] and for RDX is based on the Smith-Bharadwaj potential [19] with the modifications from Ref. [36]. Van der Waals interactions between the two force fields following the combination rules described in the Dreiding paper [20]. The particle-particle particle-mesh (PPPM) method [39] is used for long-range electrostatic interactions. Our in-house composite builder PBXgen is described in detail in Ref. [18]. LAMMPs input files in electronic form are included in the SI.

Microstructural parameters and hotspot temperature are calculated based on the centers of mass of molecules (PS monomer is treated as a molecule). Each simulation describes two shocks that are separated to generate training, validation, and testing datasets. Each subsystem is further mirrored in the thickness direction as a data augmentation strategy. Each microstructure is binned into voxels (the dimension of a voxel is about 1.9 nm x 1.1 nm x 2.8 nm) with a fixed number of bins in x, y, and z directions. To make the dimensions of voxels approximately equal for all datasets, the number of bins is set to be 32x16x32 for small subsystems, and 44x16x80 for large subsystems.

The calculations of input and output fields are based on the number of molecules included in each voxel. The equations are listed in the following:

$$\rho_{total} = (N_{RDX} M_{RD} + N_{PS} M_{PS})/V_{voxel} \quad (1)$$
$$\rho_{RDX} = N_{RDX} M_{RDX}/V_{voxel} \quad (2)$$
$$\rho_{PS} = N_{PS} M_{PS}/V_{voxel} \quad (3)$$
$$\Delta_{interface} = \rho_{RDX} \rho_{PS} \quad (4)$$
$$T = \sum T_{mol}/N_{mol} \quad (5)$$

Where $\rho$ stands for mass density, N represents the number of molecules in the bin, M molecular mass, V the volume. Subscripts indicate the molecule type. $T_{mol}$ represents the instantaneous intramolecular temperature of each molecule measured from their center of mass.

The three input fields are based on the initial PBX structure. But the output field represents the shock temperature approximately 1 ps after the passage of the shock. This requires tracking the shock front and building the temperature field one slice at a time. This is done since heat dissipation affects the temperature field and computing the temperature field for the entire system at a given time would result in different physics at different locations.

**Machine learning models (MISTnet)**

MISTnet maps the initial microstructure to shock temperature through convolutional and pooling layers down to a latent dimension, and convolution and upsampling layers back to the original 3D volume dimension. Figure 6 shows the details of our architecture. Each miniblock of the encoding stage includes two convolutional operations with a 3x3x3 kernel, and a rectified linear (ReLU) activation, a single 2x2x2 max pooling layer with stride 2. Each encoding miniblock doubles the feature maps while reducing the spatial dimensions by half.

Each miniblock of the decoding stage merges the result of a transposed convolutional operation with kernel 3x3x3 from the previous layer and the "skip connection" layer equivalent from the encoding stage. Two more convolutional operations with kernel 3x3x3 are added afterward, also followed by a linear (ReLU) activation. At the final layer of the model, a 1x1x1 convolutional operation returns the feature map to a single channel with the starting spatial dimensions. The dimensions of our input systems are 32x 16 x 32 x 3, with the first three dimensions representing the spatial coordinates last representing the number of channels.

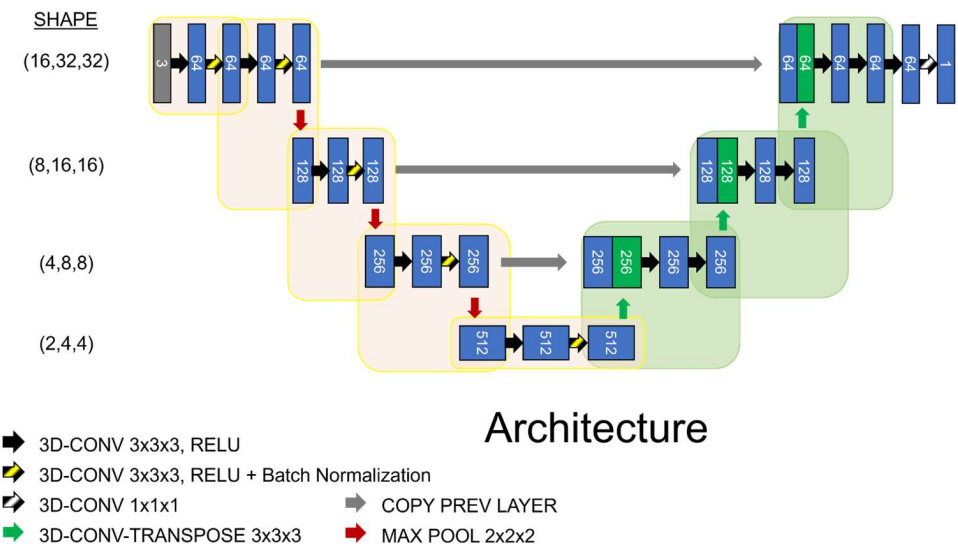

**Figure 6**. Schematic representation of the MISTnet framework architecture based on U-Net [17]. Arrows represent different operations in between layers. Dimensions of our feature maps are listed per layer, and quantity of these maps is overlayed in each layer.

## ACKNOWLEDGEMENTS

This research was sponsored by the Army Research Laboratory and was accomplished under Cooperative Agreement Number W911NF-20-2-0189.  This work was supported in part by high-performance computer time and resources from the DoD High Performance Computing Modernization Program.  The views and conclusions contained in this document are those of the authors and should not be interpreted as representing the official policies, either expressed or implied, of the Army Research Laboratory, or the U.S. Government. The U.S. Government is authorized to reproduce and distribute reprints for Government purposes notwithstanding any copyright notation herein.

## Contributions

C.L. created the atomistic models and performed and analyzed the MD simulations. J.C.V. designed, created, trained, and tested the MISTnet models. B.H.L. performed MD simulations and analyzed MD and MISTnet results, and also provided a support in refining the training procedure for MistNET. R.J.A. quantified hotspots in MD and MISTnet results. A.S. conceived the project and drafted the manuscript. All authors contributed to writing the manuscript.

## Data availability

Data used to train and test the MISTnet model is available along with the source code at https://github.com/Jverduzc/MISTnet. Data includes all datasets from MD simulations after binning. An interactive HTML file per system is also included to explore the accuracy of the predictions.

## Code availability

Source code is available at https://github.com/Jverduzc/MISTnet. A tool with an interactive notebook to test custom systems with a trained MISTnet model is available at https://nanohub.org/tools/mistnet/